\documentclass[showpacs,showkeys,amsmath,amssymb,superscriptaddress,reprint, preprintnumbers,nofootinbib,notitlepage, groupdaddress,showabstract,aps,prl,twocolumn]{revtex4-1}

\pdfoutput=1 % if your are submitting a pdflatex (i.e. if you have
\usepackage[font=small]{caption}
\usepackage[utf8]{inputenc}
\usepackage[english]{babel}
\usepackage{bm}
\usepackage{graphicx}
\usepackage{dcolumn}
\usepackage{pbox}
\usepackage{epsfig}
\usepackage[dvipsnames]{xcolor}
\usepackage{slashed}
\usepackage{amssymb}
\usepackage{ mathrsfs }
\usepackage{color}
\usepackage[font=small]{caption}
\usepackage[font=small]{subcaption}
\usepackage{url}
\definecolor{DarkBlue}{rgb}{0.7, 0.4, 1} 
\definecolor{Blue}{rgb}{0, 0.8, 0} 
\definecolor{MyLightBlue}{rgb}{0.5,0.7,1.9}
\definecolor{MyGreen}{rgb}{0.0,0.2, 0.0}
\definecolor{MyBrickRed}{rgb}{0, 0.5, 0.2}
\RequirePackage{hyperref}
\hypersetup{colorlinks, citecolor=Blue,linkcolor=DarkBlue, urlcolor=Green}
\usepackage[normalem]{ulem}
\DeclareUnicodeCharacter{2212}{-}
\DeclareUnicodeCharacter{2032}{'}
\newcommand{\bea}{\begin{eqnarray}}
\newcommand{\eea}{\end{eqnarray}}

\makeatletter

\renewcommand\@makecaption[2]{%
  \par
  \vskip\abovecaptionskip
  \begingroup
  
   \small\rmfamily
    \begingroup
     \samepage
     \flushing
     \let\footnote\@footnotemark@gobble
     \@make@capt@title{#1}{#2}\par
    \endgroup
  \endgroup
  \vskip\belowcaptionskip
}
\makeatother

\DeclareUnicodeCharacter{2212}{-}
\newcommand{\Mzp}{M_{Z^\prime}}

\newcommand{\gs}{g_\star}
\newcommand{\gss}{g_{\star s}}

\begin{document}
%%%%%%%%%%%%%%%%%%%%%%%%%%%%%%
\title{New constraints on $Z^\prime$ from captured dark matter annihilation in astrophysical objects} 
%%%%%%%%%%%%%%%%%%%%%
\author{Basabendu Barman}
\email{basabendu.b@srmap.edu.in}
\affiliation{Department of Physics, School of Engineering and Sciences, SRM University-AP, Amaravati 522240, India}
\author{Pooja Bhattacharjee}
\email{pooja.bhattacharjee@ung.si}
\affiliation{Center for Astrophysics and Cosmology, University of Nova Gorica, Vipavska 13, SI-5000
Nova Gorica, Slovenia}
\author{Arindam Das}
\email{arindamdas@oia.hokudai.ac.jp}
\affiliation{Institute for the Advancement of Higher Education, Hokkaido University, Sapporo 060-0817, Japan}
\affiliation{Department of Physics, Hokkaido University, Sapporo 060-0810, Japan}
%%%%%%%%%%%%%%%%%%%%%%%%%%%
\begin{abstract}   
Considering dark matter capture in astrophysical objects such as neutron stars and brown dwarfs, followed by their annihilation into two neutrino and four neutrino final states, we derive new constraints on the mass and coupling of a novel abelian gauge boson $Z^\prime$ arising from an anomaly-free $U(1)$ extension of the Standard Model.   We further confront these astrophysical limits with complementary bounds from the Planck-observed relic abundance via the freeze-in mechanism, big bang nucleosynthesis (BBN), gravitational wave signatures from cosmic strings, and searches at energy and intensity frontier experiments.
\end{abstract}
%%%%%%%%%%%%%%%%%%%%%%%%%%%
\maketitle
%%%%%%%%%%%%%%%%%%%%%%%%%%%%%%%%%%%%%%%%%%%%%%%%%%
\noindent
{\textbf{Introduction}.--}
%%%%%%%%%%%%
Weakly Interacting Massive Particles (WIMPs) have long been regarded as leading Dark Matter (DM) candidates, since they naturally account for the observed relic abundance through thermal freeze-out~\cite{Jungman:1995df,Roszkowski:2017nbc,Arcadi:2017kky,Arcadi:2024ukq}, in agreement with WMAP~\cite{WMAP:2008ydk} and Planck~\cite{Planck:2015fie} measurements. Yet, the increasingly stringent bounds on DM interaction cross sections have made their detection in direct and indirect searches progressively more elusive. This has shifted attention toward the freeze-in mechanism of DM production~\cite{Hall:2009bx}, where non-thermal, out-of-equilibrium, feebly interacting particles emerge as viable alternatives. The hallmark of this scenario is the extremely weak coupling between DM and the Standard Model (SM), ensuring that DM never attains thermal equilibrium with the visible sector. While such tiny interactions may at first suggest that frozen-in DM could only be probed gravitationally, these candidates can, in fact, be tested in a variety of terrestrial and astrophysical experiments~\cite{Bernal:2017kxu}. 

Another intriguing mechanism is DM capture, whereby celestial objects such as galaxies, stars, or planets can gravitationally attract and confine DM particles traversing through space. In this process, DM particles are drawn into the object’s strong gravitational potential and subsequently lose energy, typically assumed to occur through scattering with the SM. Once their velocity falls below the escape velocity, they become gravitationally bound, leading to an accumulation of DM within the object. The efficiency of this capture depends on both the properties of the astrophysical body and the DM mass, which must be sufficient to prevent evaporation~\cite{Garani:2021feo}, thereby constraining sensitivities in direct and indirect searches. In scenarios where DM undergoes annihilation, the captured population may produce observable secondary signals such as neutrinos, photons, or electrons~\cite{Cirelli:2005gh,
Blennow:2007tw,
Bell:2011sn,Elor:2015bho,
Leane:2023woh}. In this context, compact astrophysical systems like neutron stars (NSs) are extensively studied in the literature~\cite{Goldman:1989nd,Gould:1989gw,Kouvaris:2007ay,McDermott:2011jp,Guver:2012ba,Bramante:2013hn,Bell:2013xk,Cermeno:2016olb,Leane:2017vag,Bhattacharjee:2024pis,Bose:2021yhz, 
Nguyen:2022zwb,Bell:2023ysh}
, while brown dwarfs (BDs), which occupy the mass range between giant planets and low-mass stars and offer complementary environments~\cite{1963ApJ...137.1126K,1963ApJ...137.1121K,10.1143/PTP.30.460,Nakajima:1995sv,Rebolo1995,Berger:2001rf,Rutledge:2000nu,Leane:2020wob,Leane:2021ihh,Bhattacharjee:2022lts}, are of particular interest. In such systems, captured DM may annihilate and produce highly energetic neutrinos that could be detected by terrestrial observatories such as IceCube (IC1)~\cite{IceCube:2019cia,IceCube:2021oqo}, IceCube-Gen2 (IC2)~\cite{IceCube-Gen2:2021tmd}, KM3NeT (KM)~\cite{KM3Net:2016zxf}, and Trident (TD)~\cite{TRIDENT:2022hql}, etc.

Beyond the detection of high-energy neutrino events, observations of gravitational waves (GWs)\cite{LIGOScientific:2021nrg,Caldwell:2022qsj,NANOGrav:2023gor,NANOGrav:2023hvm,Xu:2023wog,EPTA:2023fyk,LISACosmologyWorkingGroup:2022jok} have had a profound impact on exploring physics beyond the Standard Model (BSM). In the early Universe, at extremely high temperatures, spontaneous symmetry breaking can give rise to topological defect networks such as cosmic strings and domain walls~\cite{Nielsen:1973cs,Kibble:1976sj}. These structures are capable of sourcing a prominent stochastic GW background, thereby connecting the underlying symmetry-breaking scale to new physics.

\begin{figure*}
\centering    
\includegraphics[width=1.0\textwidth,angle=0]{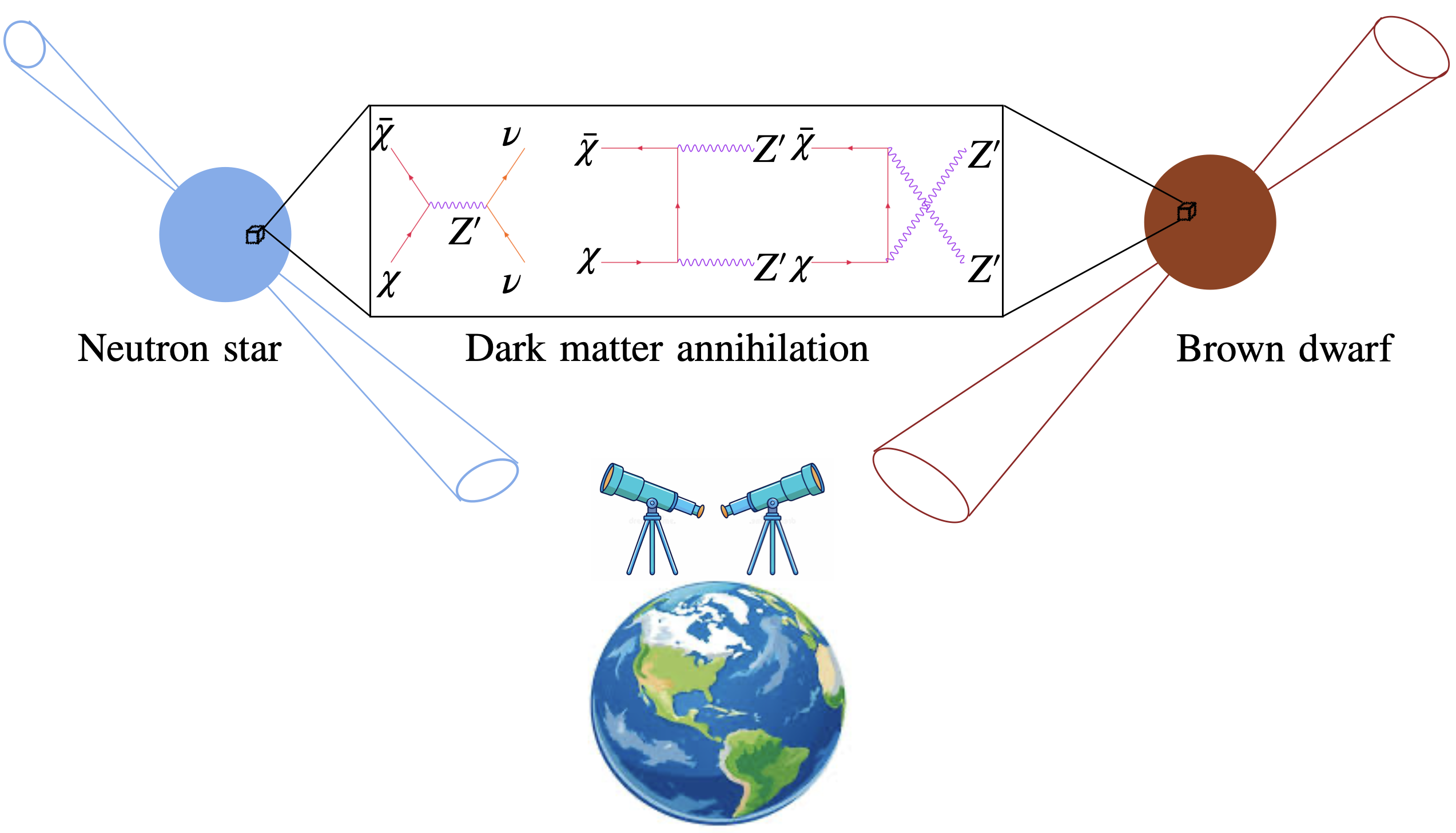}
\caption{$2\nu$ and $4\nu$ final states from distant exotic astrophysical objects like Neutron Star and Brown Dwarf. The $2\nu$ process $\chi \bar{\chi} \to \nu \bar{\nu}$ is mediated by short-lived $Z^\prime$, while $4\nu$'s are produced from $\chi \bar{\chi} \to Z^\prime Z^\prime$, followed by long-lived  $Z^\prime$ decay $Z^\prime \to 2 \nu$.}
\label{fig:summary}
\end{figure*}

In this work, we investigate the scenario where captured DM candidates annihilate inside astrophysical objects such as NSs and BDs, producing high-energy neutrinos. These neutrinos can be probed by current and future generation terrestrial detectors, such as IC1, KM3 and prospective IC2 and TD, where they may appear as two- or four-neutrino events. To realize this framework, we adopt a UV-complete, minimal extension of the SM featuring an anomaly-free $U(1)_X$ gauge symmetry~\cite{Oda:2015gna,Das:2016zue}, augmented by three generations of SM-singlet right-handed neutrinos (RHNs) and a singlet scalar field that spontaneously breaks $U(1)_X$ via a nonzero vacuum expectation value (VEV). This breaking generates Majorana masses for the RHNs, which subsequently induce tiny neutrino masses and flavor mixing through the seesaw mechanism~\cite{Minkowski:1977sc,Yanagida:1979as,Gell-Mann:1979vob,Mohapatra:1979ia,Schechter:1980gr}. 
Within this setup, we introduce a Dirac-type DM candidate that can be gravitationally captured by NSs and BDs, annihilating via the $U(1)_X$ gauge boson, $Z^\prime$, into neutrinos (see Fig.~\ref{fig:summary}). 
% We consider two scenarios after DM annihilation, where $2\nu$ final state is produced from short-lived mediator following $\chi\chi \to \nu \nu$ process assuming maximum survival probability and $4\nu$ final state via $\chi \chi \to Z^\prime Z^\prime$, $Z^\prime \to 2\nu$ where $Z^\prime$ is long-lived where survival probability depends on boost factor, the radius of the astrophysical object and its distance from the Earth. 
We analyze two post-annihilation scenarios for DM. In the first, a $2\nu$ final state is produced via $\chi\chi\to\nu\nu$ through a short-lived mediator, assuming a maximal survival probability. In the second, a $4\nu$ final state emerges from $\chi\chi\to Z'Z'$ followed by $Z'\to\nu\nu$, where the long-lived $Z'$ mediator has a survival probability determined by its boost factor, the source radius, and its distance from the Earth. By comparing the resulting flux with the all-sky projected sensitivity expected from IC1, KM3 and prospective IC2 and TD, we derive new projected bounds on the model parameter space $[g_X,\,\Mzp]$. It is important to note that we focus on old, cold NSs with core temperatures of $T_{\rm core} \sim 0.1\,\mathrm{keV}$, essentially lacking a thermal neutrino bath unlike  proto-NSs where $T_{\rm core}$ reaches a few MeV. Under such conditions, Pauli blocking and degeneracy effects suppress scattering, thereby increasing the neutrino mean free path. 
For old and cold BDs, with $T_{\rm core} \simeq 10^3$ K and densities of  200-700~g\,cm$^{-3}$, the neutrino mean free path is even larger, making trapping and heating effects negligible. Thus, neutrinos from DM annihilation are expected to escape freely in both cases. Crucially, we investigate the regime of feeble DM–SM interactions necessary for freeze-in production of the observed relic abundance, and emphasize complementary probes from GW different observatories. Additionally we show the complementarity with the bounds from existing beam-dump and low-energy scattering experiments.
\\
%%%%%%%%%%%%%%%%%%%%%%%%%
%%%%%%%%%%%%%%%%%%%%%%%%%%%%%%%%

%%%%%%%%%%%%%%%%%%%
\noindent
{\textbf{The framework}--}
We introduce a general $U(1)_X$ gauge symmetry as a linear combination of the anomaly-free 
$U(1)_{\rm Y}$ and $U(1)_{\rm B-L}$ groups, $Q_X = x_H Q_{\rm Y} + x_\Phi Q_{\rm B-L}$. Under 
$SU(3)_c \times SU(2)_L \times U(1)_Y \times U(1)_X$ the SM quarks transform as $q_L^i=\{3,2,\frac{1}{6}, x_q=\frac{1}{6} x_H+\frac{1}{3} x_\Phi\}$, $u_R=\{3,1,\frac{2}{3}, x_u=\frac{2}{3} x_H+\frac{1}{3} x_\Phi\}$, $d_R^i=\{3,1,-\frac{1}{3}, x_d=-\frac{1}{3}x_H+\frac{1}{3} x_\Phi\}$, respectively. The SM leptons transform as $\ell_L^i=\{1,2,-\frac{1}{2}, x_\ell=-\frac{1}{2}x_H-x_\Phi\}$, $e_R^i=\{1,1,-1, x_e=-x_H-x_\Phi\}$ where $i$ stands for three generations of the fermions. Here SM-like Higgs field is transformed as $H=\{1,2,\frac{1}{2}, x_h=\frac{1}{2}x_H\}$ and we introduce an SM-singlet $U(1)_X$ scalar, responsible for the $U(1)_X$ symmetry breaking, transforms following $\Phi=\{1,1,0, x_\Phi\}$. Three generations of RHNs $N_R^i$ are introduced to realize the seesaw mechanism~\cite{Oda:2015gna,Das:2016zue}. Without loss of generality, we set $x_\Phi = 1$. Specific choices of $x_H$ recover well-known scenarios, such as,
\begin{align*}
& x_H=
\begin{cases}
0\,, & U(1)_{\rm B-L}
\\[10pt]
-1\,, & Q_X(e_R)=0 \; (Q_X(d_R)=0)
\\[10pt]
-2\,, & Q_X(q_L,\ell_L)=0 \; \; (U(1)_R)
\\[10pt]
+2\,, & \text{purely chiral}~U(1)_X\,.
\end{cases}
\end{align*}

The relevant part of the Lagrangian manifesting the neutrino mass generation can be written as 
\bea
\mathcal{L}_{\rm{Yukawa}}\supset -Y_{\nu_{ij}} \overline{\ell_L^i} \widetilde{H} N_R^j -\frac{1}{2} Y_{N_\alpha} \overline{\left({N_R^\alpha}\right)^c}\,N_R^\alpha\,\Phi+\text{h.c.}
\label{eq:LYk}
\eea
The scalar potential involving two scalar fields is given by 
\bea
V=\sum_{\mathcal{I}= H, \Phi} \Big[m_{\mathcal{I}}^2 (\mathcal{I}^{\dagger} \mathcal{I})+ \lambda_{\mathcal{I}} (\mathcal{I}^\dagger \mathcal{I})^2 \Big] +
\lambda^\prime (H^\dagger H) (\Phi^{\dagger} \Phi)\,.
\label{pot}
\eea 
After $U(1)_X$ and electroweak gauge symmetries are broken, the scalar fields $H$ and $\Phi$ develop their respective VEVs as 
\begin{align}\label{eq:VEV}
  \langle H\rangle \ = \ \frac{1}{\sqrt{2}}\begin{pmatrix} v+h\\0 
  \end{pmatrix}~, \quad {\rm and}\quad 
  \langle\Phi\rangle \ =\  \frac{v_\Phi^{}+\phi}{\sqrt{2}}~,
\end{align}
where electroweak scale is $v=246$ GeV at the potential minimum. 
Considering $v_\Phi\gg v$, we obtain the mass of the $U(1)_X$ gauge boson as $M_{Z^\prime}^{}= 2 g_X^{}  v_\Phi^{}$. For simplicity we consider very small mixing\footnote{Bounds from LHC~\cite{Robens:2015gla,Chalons:2016jeu,Das:2022oyx}, LEP~\cite{LEPWorkingGroupforHiggsbosonsearches:2003ing}, prospective colliders like ILC~\cite{Wang:2020lkq} and CLIC~\cite{CLIC:2018fvx} indicate that the scalar mixing could be $<0.001$, for $\phi$ mass up to 1 TeV.} between the physical states of the scalars, as well as between $Z$ and $Z^\prime$ \footnote{The upper bound on the kinetic mixing parameter could be $<10^{-2}$~\cite{Leike:1992uf,Carena:2004xs}, making kinetic-mixing induced scattering processes inefficient.}. The Majorana mass of the RHNs $M_{\alpha}= Y_{N_\alpha} v_\Phi/\sqrt{2}$ and the Dirac mass of the light left-handed neutrinos $m_{{D}_{\alpha \beta}}= Y_{{\nu}_{\alpha \beta}}v /\sqrt{2}$ are generated after the breaking of $U(1)_X$ and electroweak gauge symmetries from Eq.~\eqref{eq:LYk}. Hence light active neutrino masses can be induced through the standard seesaw formula $m_\nu\sim -m_D^{} M_\alpha^{-1} m_D^T$~\cite{Gell-Mann:1979vob,Sawada:1979dis, Mohapatra:1980yp} successfully explaining the origin of tiny neutrino mass and flavor mixing. 

Finally, as a viable DM candidate, we introduce gauge singlet $(\chi_L, \chi_R)$ having $U(1)_X$ charge $Q_\chi$ and interacting with the $Z^\prime$ boson via 
\begin{align}
\mathcal{L}_{\rm DM}= i \overline{\chi} \gamma^{\mu} (\partial_{\mu}+ i g_X Q_{\chi} Z^\prime_{\mu}) \chi + m_{\chi} \overline{\chi} \chi\,,
\end{align}
where $\chi=\chi_L+\chi_R$. To prevent the decays of $\chi_{L, R}$ into the RHNs and other particles we prohibit $Q_\chi=\pm x_\Phi, \pm 3 x_\Phi$ to ensure stability of the DM candidate. In the rest of the analysis we consider $Q_\chi=100$ and in addition to that we consider 
$m_\chi > 2 M_{Z^\prime}$ to forbid the decay of $Z^\prime$ into $\chi$. 
\\ \\
%%%%%%%%%%%%%%%%%%%%%%%%%%%%%%%%%%%%%%%%%%%%%%%%%%%%%%%
\noindent
{\textbf{DM-capture by astrophysical objects}--} We explore two scenarios where DM annihilates into (i) light neutrinos and (ii) long-lived mediators which decay into light neutrinos outside these systems. These mechanisms are naturally motivated within secluded DM models predicting such mediators~\cite{Pospelov:2007mp, Pospelov:2008jd, Gherghetta:2015ysa}. As the DM from the Galactic halo transits through the celestial objects, it can scatter with the stellar material and lose energy and eventually once the kinetic energy of the DM is less than the gravitational potential, the DM will be captured. The expression for maximum DM capture rate \( C_{\rm max} \) by our targets is given by~\cite{Leane:2021ihh,Garani:2017jcj}~\footnote{Here we assume that all DM that passes through the effective area of the BD/NS is captured and neglect the effect of the motion of the compact object with respect to the DM halo.},
\begin{equation}
C_{\rm{max}}(r) = \pi R_{\star}^{2} n_{\chi}(r) v_{0} \left( 1+ \frac{3}{2} \frac{v_{\rm{esc}}^2}{\bar{v}(r)^2} \right), \label{eqn:Cmax}
\end{equation}
with DM number density $n_{\chi}=\rho_\chi/m_\chi$~\cite{Calore:2022stf}, \( v_{0} = \sqrt{8/(3\pi)}\,\bar{v} \), \( \bar{v}(r) = 3\,v_c(r)/2 \), and \( v_c(r) = \sqrt{G_N\,M(r)/r} \). We adopt Navarro--Frenk--White (NFW) profile~\cite{Navarro:1995iw,Navarro:1996gj}
for the DM distribution including gravitational focusing via \( v_{\rm esc} =  \sqrt{2\,G_N\,M_\star/R_\star} \), where $R_\star$ and $M_\star$ are radius and mass of our targets, respectively. For NS, due to blue-shifted incoming DM velocities, \( v_{\rm esc}\) becomes, \( v_{\rm esc} = \sqrt{2 \chi_c}\), where, \( \chi_c = 1- \sqrt{1-2\,G_N\,M_\star/R_\star}\) \cite{Leane:2021ihh}. Typically, DM is assumed to be captured after a single scattering within BDs/NSs, but for heavy DM this assumption fails, since multiple collisions are needed to dissipate sufficient kinetic energy for capture~\cite{Bramante:2017xlb}. The probability that DM with optical length $\eta$ undergoes $N$ actual scatterings is given by
%%%%%%%%%%%%%%%%%%%%%%%%%%%%%
\begin{equation}
p_N(\eta) = 2 \int_0^1 dy \, y\, e^{-y\eta} \frac{(y\eta)^N}{N!}, \label{eqn:Ctot}
\end{equation}
%%%%%%%%%%%%%%%%%
where $\eta = \left(3/2\right)\,\left(\sigma_{\chi n}/\sigma_{\rm sat}\right)$~\footnote{The optical length is defined such that $\eta=1$ corresponds to DM scattering once on average while traversing BDs/NSs.} is the optical length, with $\sigma_{\chi n}$ denoting the DM-nucleon scattering cross section, $\sigma_{\rm sat} = \pi R_{\star}^2 / N_n$ is the saturation cross-section, and number of nucleons in the target $ N_n = M_{\star}/m_n $, where $m_n$ is the nucleon mass.
%%%%%%%%%%%%%%%%%%%%%%%%%%%%%%%%%%%%%%%%
\par The total capture rate, accounting for multiple $N$ scatterings is then given by~\footnote{In practice, the sum can be truncated at some finite maximum $N \gg \eta$, since the probability of scattering more than $\eta$ times is exponentially suppressed.},
\begin{equation}
C_{\rm tot}(r) = \sum_{N=1}^{\infty} C_N(r), \label{eqn:Ctot}
\end{equation}
where $C_N$, defined below, denotes the capture rate for particles that undergo $N$ scatterings.
Assuming constant \( v_{\rm esc} \) and \( N_n \) across our targets, the $N$-scattering rate is given by
\begin{align}
C_{N}(r) &= \frac{\pi R_{\star}^{2} p_{N}(\eta)}{1 - \frac{2G_N M_{\star}}{R_{\star}}} 
\frac{\sqrt{6}\, n_{\chi}(r)}{3\sqrt{\pi} \bar{v}(r)} \nonumber \\
&\quad \times \left[ (2\bar{v}(r)^{2} + 3v_{\rm{esc}}^{2}) 
- (2\bar{v}(r)^{2} + 3v_{N}^{2}) \right. \nonumber \\
&\quad \left. \times \exp\left(-\frac{3(v_{N}^{2} - v_{\rm{esc}}^{2})}{2\bar{v}(r)^{2}} \right) \right]\,,
\label{eq:C_N}
\end{align}
where \( v_N = v_{\rm esc}(1 - \beta_+/2)^{-N/2} \), \( \beta_+ = \frac{4m_\chi m_n}{(m_\chi + m_n)^2} \). For a large number of scatterings ($N \gg 1$), $C_{N}$ approximates $\approx p_{N} C_{\rm{max}}$. Under the conservative estimation, $C = \min[C_{\rm{tot}}, C_{\rm{max}}]$ denotes the total capture rate as not all DM particles passing through are captured.
%%%%%%%%%%%%%%%%%%%%%
\par In presence of DM self-annihilation, there will be an interplay between the capture and annihilation rate in the BDs/NSs. Neglecting evaporation\footnote{Evaporation happens when the system’s core is hot enough to give DM sufficient kinetic energy and has a low enough gravitational potential to allow escape~\cite{Leane:2020wob,Bell:2020lmm}.}~(relevant for $ m_\chi >m_{\rm evap}$ (evaporation mass)), the DM number evolution inside an object is~\cite{Kouvaris:2010vv}, 
\begin{equation}
\frac{dN_\chi(t)}{dt} = C - C_{\rm ann} N_\chi^2(t), \quad 
C_{\rm ann} = \frac{\langle \sigma v \rangle}{V_{\rm *}},
\end{equation}
with solution
\begin{equation}
N_\chi(t) = \sqrt{\frac{C}{C_{\rm ann}}}\, t_{\rm eq} \tanh\left(\frac{t}{t_{\rm eq}}\right), \,\,
t_{\rm eq} = \frac{1}{\sqrt{C_{\rm ann} C}}\, ,
\end{equation}
while $C_{\rm ann}$ signifies the annihilation rate, where $V_{\rm *}$ denotes the volume within the target in which annihilation occurs and depends on the core density and core temperature of the target \cite{Bhattacharjee:2022lts}. The equilibrium time, $t_{\rm eq}\equiv 1/\sqrt{C_{\rm ann}\,C}$, represents the time required to reach equilibrium between the DM capture rate and the annihilation rate in the absence of evaporation. The DM annihilation rate is $\Gamma_{\rm ann}=(1/2)\,C_{\rm ann}N_\chi(t)^2$, where the factor $1/2$ comes from DM self-annihilation.
%%%%%%%%%%%%%%%%%%%%%%%%%
\begin{figure*}
\centering    
\includegraphics[width=0.46\textwidth,angle=0]{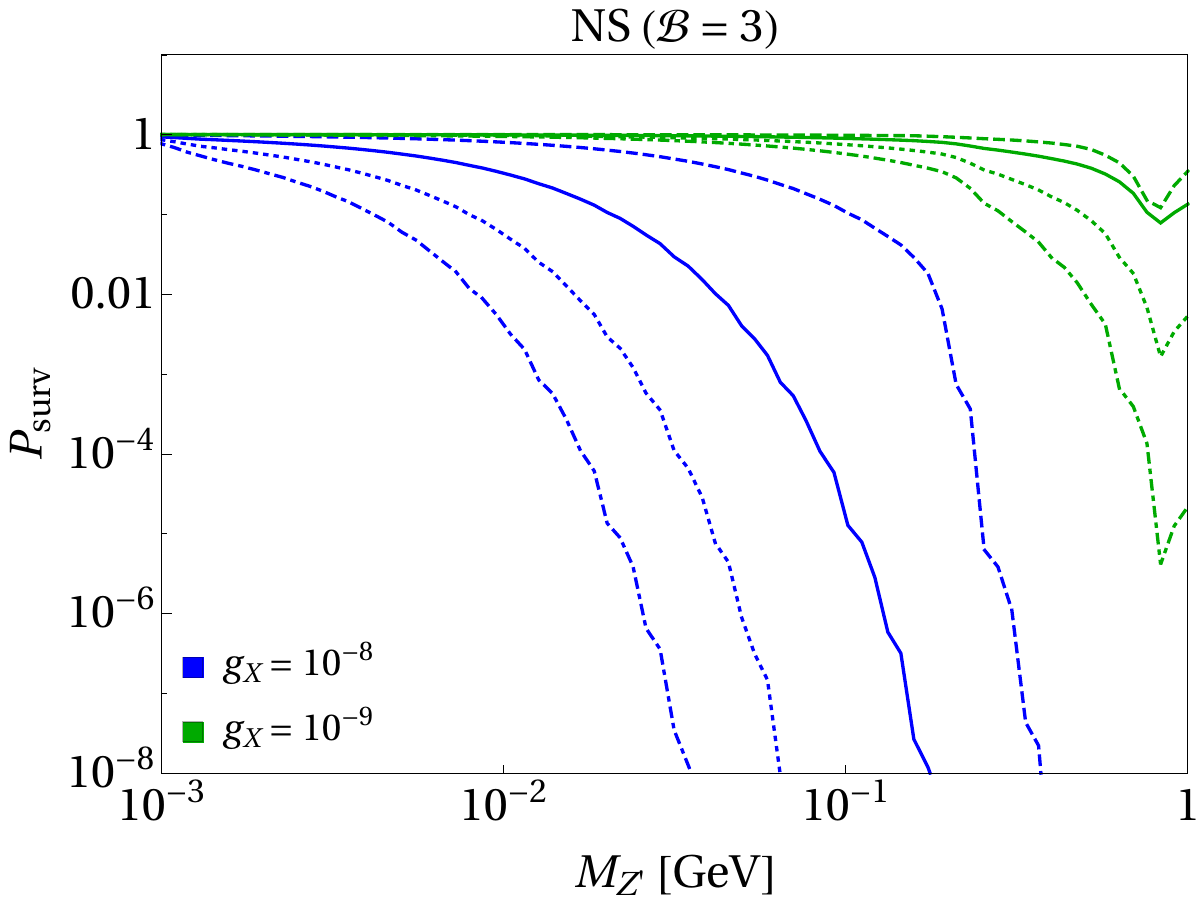}
\includegraphics[width=0.46\textwidth,angle=0]{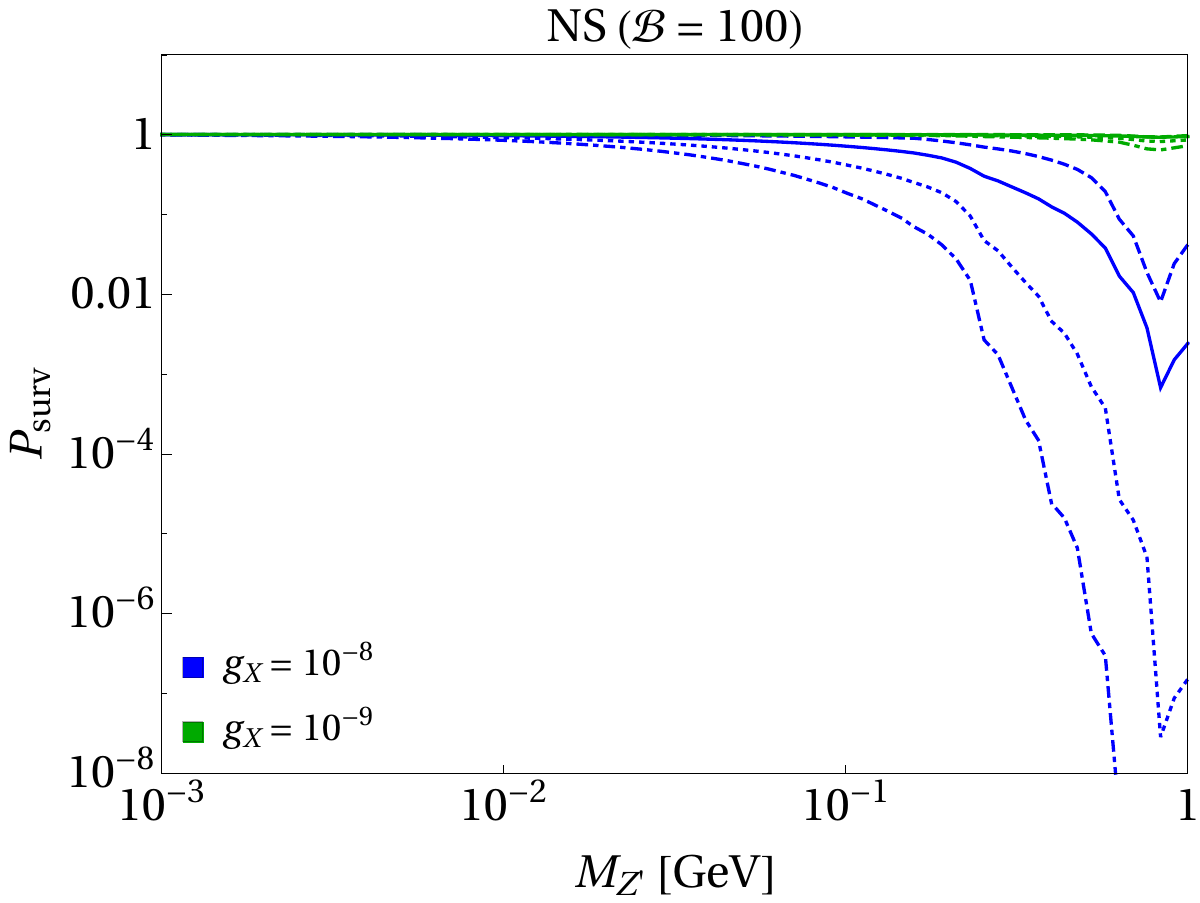}
\caption{Survival probability for NS with $\mathcal{B}=\{3,\,100\}$ in the left and in the right panel, respectively. Different charges $x_H=\{0,\,-1,\,1,\,2\}$ are shown respectively via solid, dashed, dotted and dot-dashed curves.}
\label{fig:Psurv}
\end{figure*}
%%%%%%%%%%%%%%%%%%%%%%%%%%%%%%%%%%%%%%%%%%%%%%%%%%%%

Now let us discuss the neutrino flux resulting from DM annihilation in BDs/NSs, as shown in Fig.~\ref{fig:summary}. The thermally averaged cross section $\langle \sigma v \rangle$ for $\chi \chi \to \nu \nu$ process can be written as,  
\bea
\bigl< \sigma v \bigr>_\nu = 
	 \frac{Q_\chi^2 g_X^4 }{ 2\pi m_\chi^2 }
	\frac{ \sqrt{ 1 - x_f^2 }\,(x_H/2+x_\Phi^{})^2 ( 1 - x_f^2)}{(x_{z^\prime}^2-4)^2 + x_{z^\prime}^4\,\Gamma_{Z^\prime}^2/M_{Z^\prime}^2 } ,
\eea
where $x_{f(z^\prime)}=m_f (M_{Z^\prime})/ m_\chi$ and $\Gamma_{Z'}=\Gamma_{Z'}\left(g_X,\,\Mzp,\,x_H\right)$ is the total decay width including all possible SM final states of $Z^\prime$. The thermally averaged cross section $\langle \sigma v \rangle$ for $\chi \chi \to Z^\prime Z^\prime$ process can be written as,
\bea
\bigl< \sigma v \bigr>_{Z^\prime} = 
	\frac{ Q_\chi^4 g_X^4 }{ 4 \pi m_{\chi}^2 }
	\frac{ ( 1 - x_{z^\prime{}}^2 )^{3/2} }{ ( 2 - x_{z^\prime{}}^2 )^2 }.
 \label{zz}   
\eea
The partial decay widths of $Z^\prime$ into a pair of SM fermions can be written as
\bea
\Gamma(Z' \to \bar{f} f)&=& N_C^{} \frac{M_{Z^\prime}^{} g_{X}^2}{24 \pi} \Bigg[ \left( q_{f_L^{}}^2 + q_{f_R^{}}^2 \right) \left( 1 - \frac{m_f^2}{M_{Z^\prime}^2} \right)+ \nonumber \\
&+&6 q_{f_L^{}}^{} q_{f_R^{}}^{} \frac{m_f^2}{M_{Z^\prime}^2}\Bigg] \left( 1 - 4 \frac{m_f^2}{M_{Z^\prime}^2} \right)^{\frac{1}{2}}\,,
\label{eq:width-ll}
\eea   
where $m_f^{}$ is the SM fermion mass, $q_{f_{L(R)}}$ is the $U(1)_X$ charge of the left (right) handed fermions and $N_C^{}=1 (3)$ being the color factor for the SM leptons (quarks). In this analysis we consider RHNs are heavier than $Z^\prime$. Assuming DM annihilates either directly to neutrinos following $\chi \chi \to \nu \nu$ only through $Z^\prime$ inside the celestial bodies. While considering the process $\chi \chi \to Z^\prime Z^\prime$, we consider $Z^\prime$ can decay outside the object, hence following, $\big<\sigma v \big>_{Z^\prime}\times$ 2 BR$(Z^\prime \to \nu \nu)$ where we estimate BR$(Z^\prime \to \nu \nu)$ where partial decay width of $Z^\prime$ into $2\nu$ is
\bea
\Gamma(Z^\prime \to 2\nu) = \frac{g_X^2 x_\ell^2}{24\pi} M_{Z^\prime}.
\eea
The resulting neutrino flux in this scenario is 
\begin{equation}
\frac{d\phi_{\nu}}{dE_{\nu}} = \frac{\Gamma_{\rm ann}}{4\pi d_{\rm *}^2} \left( \frac{dN_{\nu}}{dE_{\nu}} \right) P_{\rm surv},
\label{eq:exp_fluc}
\end{equation}
where $d_\star$ denotes the distance to the celestial body from the Earth and $\frac{dN_\nu}{dE_{\nu}}$ is the expected neutrino spectrum. The survival probability of neutrinos,
\bea\label{eq:Psurv}
P_{\rm surv}=e^{-R_\star/ \mathcal{B}c\tau_{Z^\prime}}-e^{-d_\star/\mathcal{B}c\tau_{Z^\prime}}, 
\eea
becomes relevant when DM annihilates into two $Z^{\prime}$ through $t-$ and $u-$ channels, each of which subsequently decays into $2\nu$ outside the celestial body and $c\tau_{Z^\prime}$ is the decay length of the $Z^\prime$. Hence we define the quantity 
\bea\label{eq:ctau}
\mathcal{B}\,c\tau_{Z^\prime}=\mathcal{B} \times
\frac{1.97\times 10^{-16}}{\Gamma_{Z'}} [\rm m]\,,
\eea
where $\mathcal{B}=m_{\chi}/M_{Z^\prime}$ is the boost factor. In our case $P_{\rm surv} \neq 0$, with the maximum flux expected for $P_{\rm surv} = 1$. For both NS and BD cases, $P_{\rm surv}$ drops sharply as $\Mzp$ increases. In BD scenario, $P_{\rm surv}=1$ can be attained for $\Mzp < 1~\mathrm{MeV}$. The variation of $P_{\rm surv}$ with respect to the model parameters is shown in Fig.~\ref{fig:Psurv}, for NS. For direct annihilation into $2\nu$, the flux  depends on many factors\footnote{In NSs, the extreme density 
($\rho \sim 10^{14}\,\mathrm{g\,cm^{-3}}$) causes the neutrino mean free path to fall below the stellar radius already at $E_\nu \gtrsim 10$~GeV, whereas in BSs, with much lower densities ($\rho \sim 10^{2}$--$10^{3}\,\mathrm{g\,cm^{-3}}$), neutrinos 
can escape up to $E_\nu \sim 100$~GeV before trapping becomes relevant. For both BD and NS, within our chosen $m_{\chi}$ range, the neutrino mean free path is expected to exceed the stellar radius, $R_{\star}$ \cite{Reddy:1997yr, Bahcall:1989ks}.}, such as, the neutrino mean free path, diffusion time, temperature and density of core, etc. As the neutrino kinetic energy strongly depends on the DM mass, for old and cold systems, they can escape the objects and potentially yield a detectable flux in neutrino telescopes. Depending on the annihilation channel producing $2\nu$ from short-lived $Z^\prime$, the spectral shape becomes Dirac-delta function 
\bea
\left( \frac{dN_{\nu}}{dE_{\nu}} \right)= \delta(E_\nu-m_\chi)
\eea
and the spectrum becomes box-shaped for $4\nu$ final state from two long-lived $Z^\prime$
\bea
\left( \frac{dN_{\nu}}{dE_{\nu}} \right)= \frac{4}{\Delta E}\Theta(E_\nu-E_-)\Theta(E_\nu-E_+).
\eea
where $\Theta$ is the step function, $E_{\pm}=\frac{1}{2} (m_\chi \pm \sqrt{m_\chi^2-M_{Z^\prime}^2})$ and $\Delta E=\sqrt{m_\chi^2-M_{Z^\prime}^2}$ \cite{ Ibarra:2012dw,Dasgupta:2012bd}. The $2\nu$ channel provides the strongest constraints from neutrino telescopes, with the delta-like spectrum modeled as a Gaussian (smeared by detector resolution) and arriving at Earth in a $1:1:1$ flavor ratio due to oscillations. In this section, we present the framework for DM capture and annihilation in astrophysical targets, incorporating target-specific effects such as Pauli blocking for $m_{\chi} < 1$ GeV in NSs, which significantly affect the expected signals \cite{Bhattacharjee:2023qfi, Bhattacharjee:2022lts}.
\\

%%%%%%%%%%%%%%%%%%%%%%%%%
\begin{figure*}
\centering    
\includegraphics[width=0.4973\textwidth,angle=0]{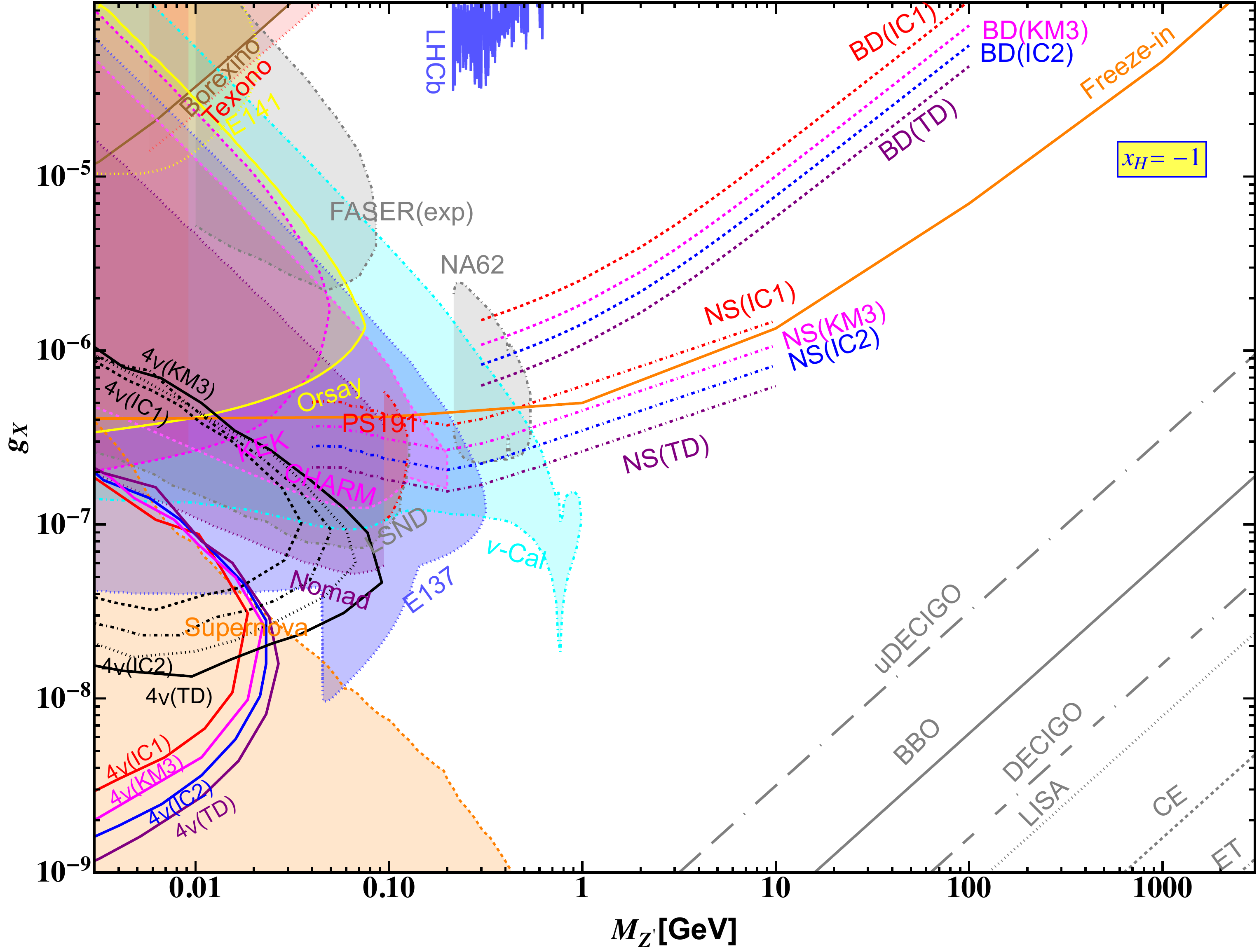}
\includegraphics[width=0.4973\textwidth,angle=0]{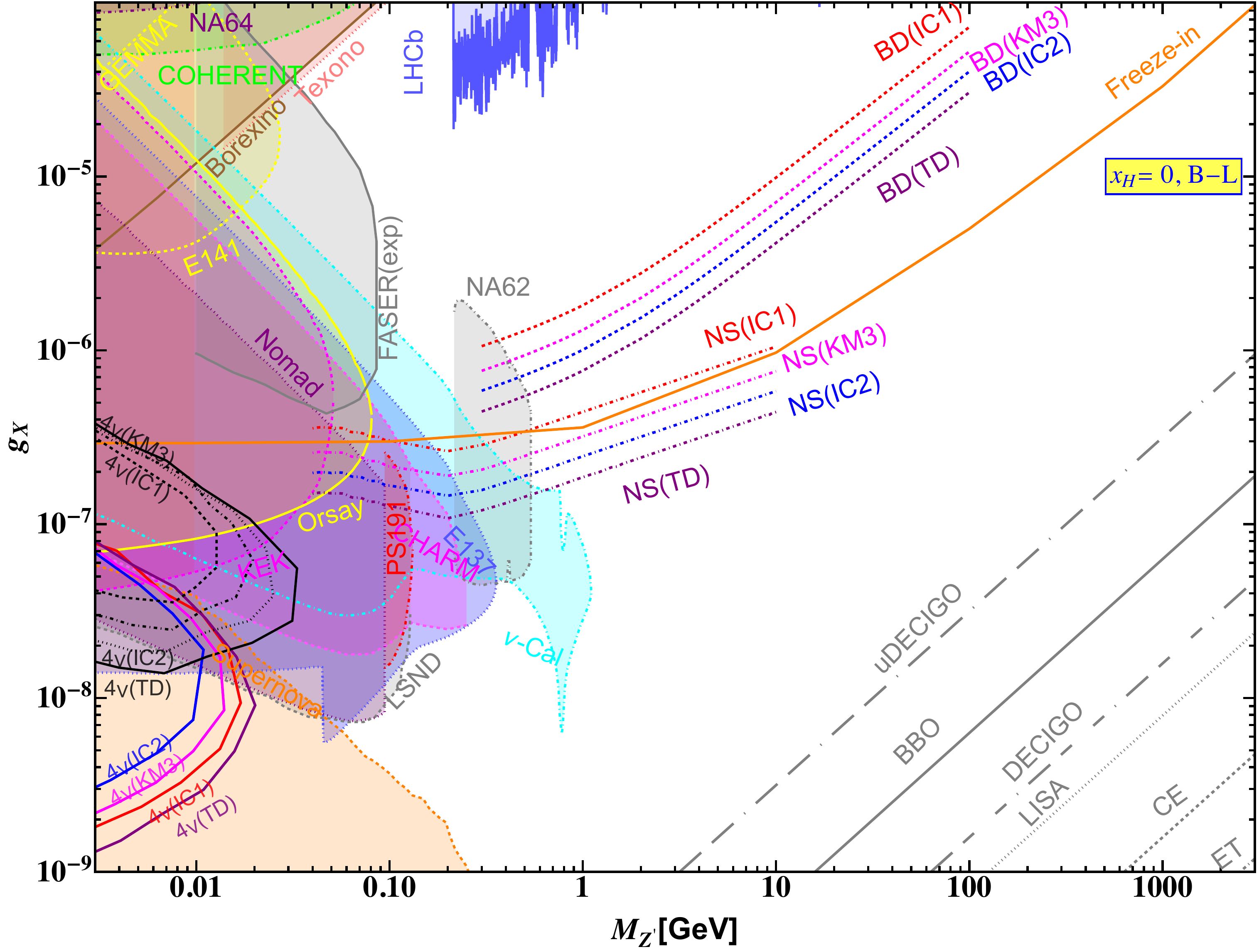}
\includegraphics[width=0.4973\textwidth,angle=0]{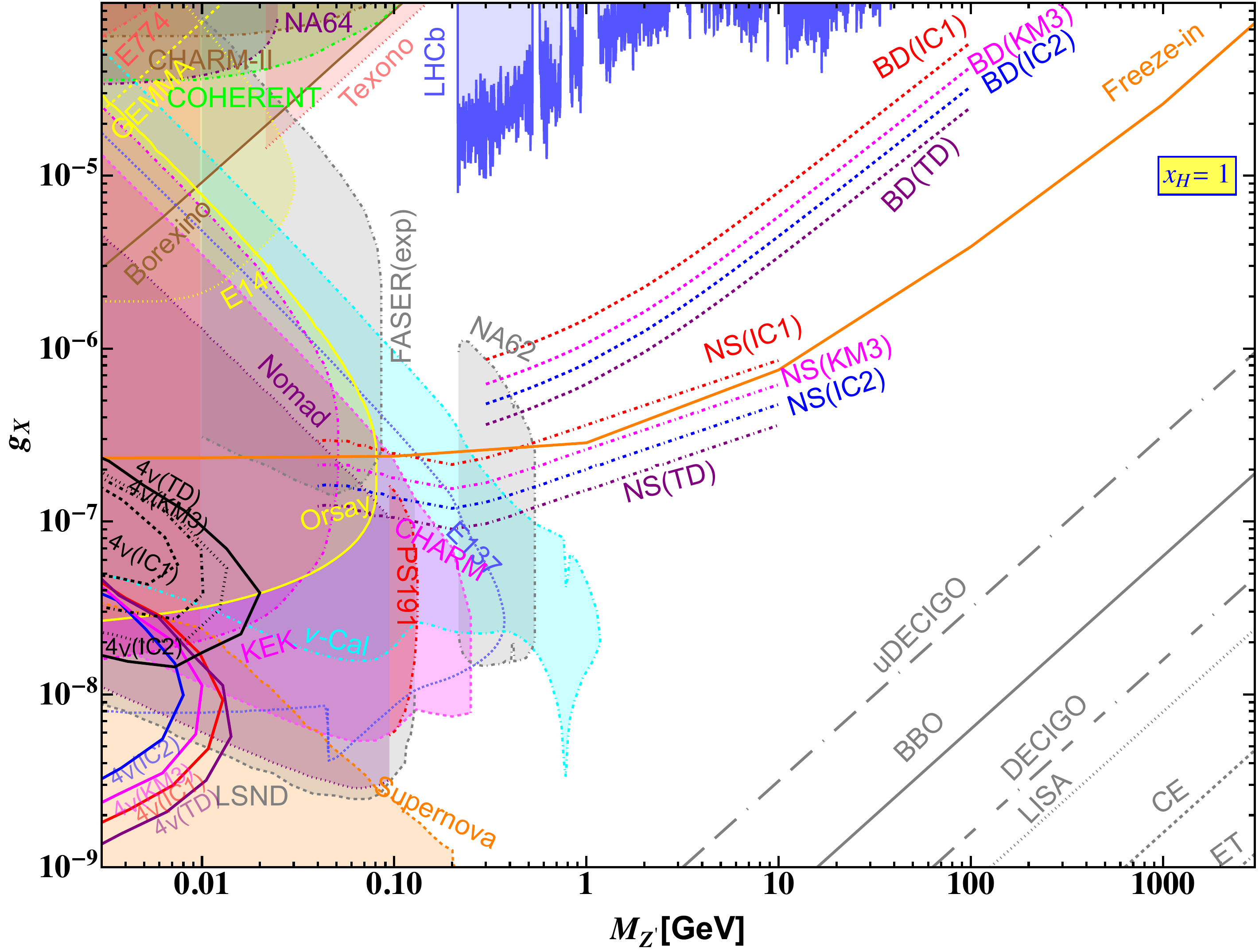}
\includegraphics[width=0.4973\textwidth,angle=0]{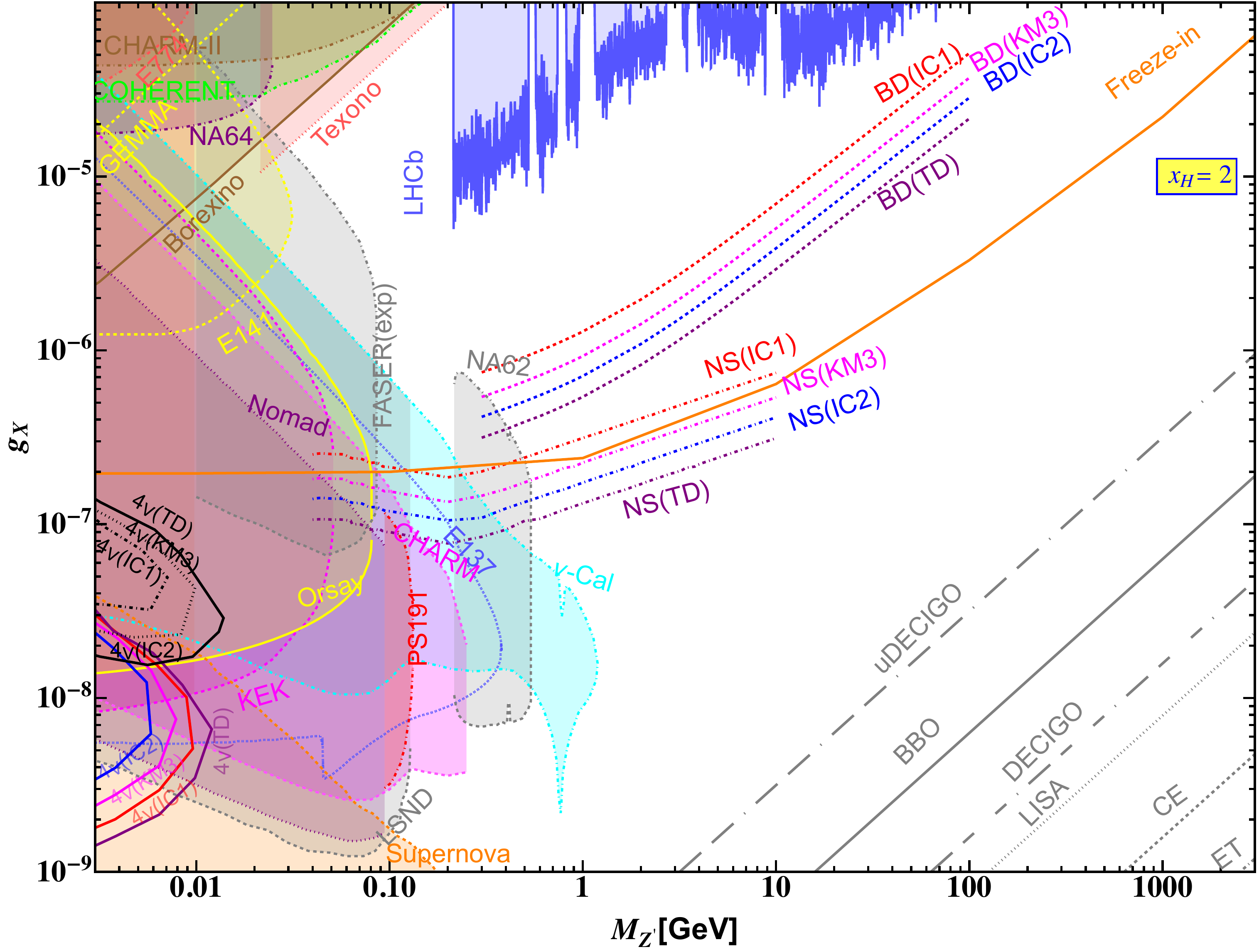}
\caption{Limits on $g_X-M_{Z^\prime}$ plane from DM annihilation comparing with $2\nu$ (NS and BD) considering short-lived $Z^\prime$ and $4\nu$ (NS) final states considering long-lived $Z^\prime$ for different $x_H$ considering $Q_\chi=100$ $m_\chi=3\,\Mzp$. Black contours for $4\nu$ final state considering $m_\chi= 100 M_{Z^\prime}$ using long-lived $Z^\prime$ from NS. We show contour of right relic abundance for DM production from freeze-in process via $s$-channel $Z'$-mediated process. Gray diagonal lines represent prospective limits from GW search experiment (uDECIGO, BBO, DECIGO, LISA, CE, ET) where GW is considered to be produced from cosmic strings. Shaded regions are ruled out by existing experimental bounds.}
\label{fig:summary1}
\end{figure*}
%%%%%%%%%%%%%%%%%%%%%%%%%%%%%%%%%%%%%%%%%%%%%%%%%%%%%%%%%%%%
\begin{figure*}
\centering    
\includegraphics[width=0.4973\textwidth,angle=0]{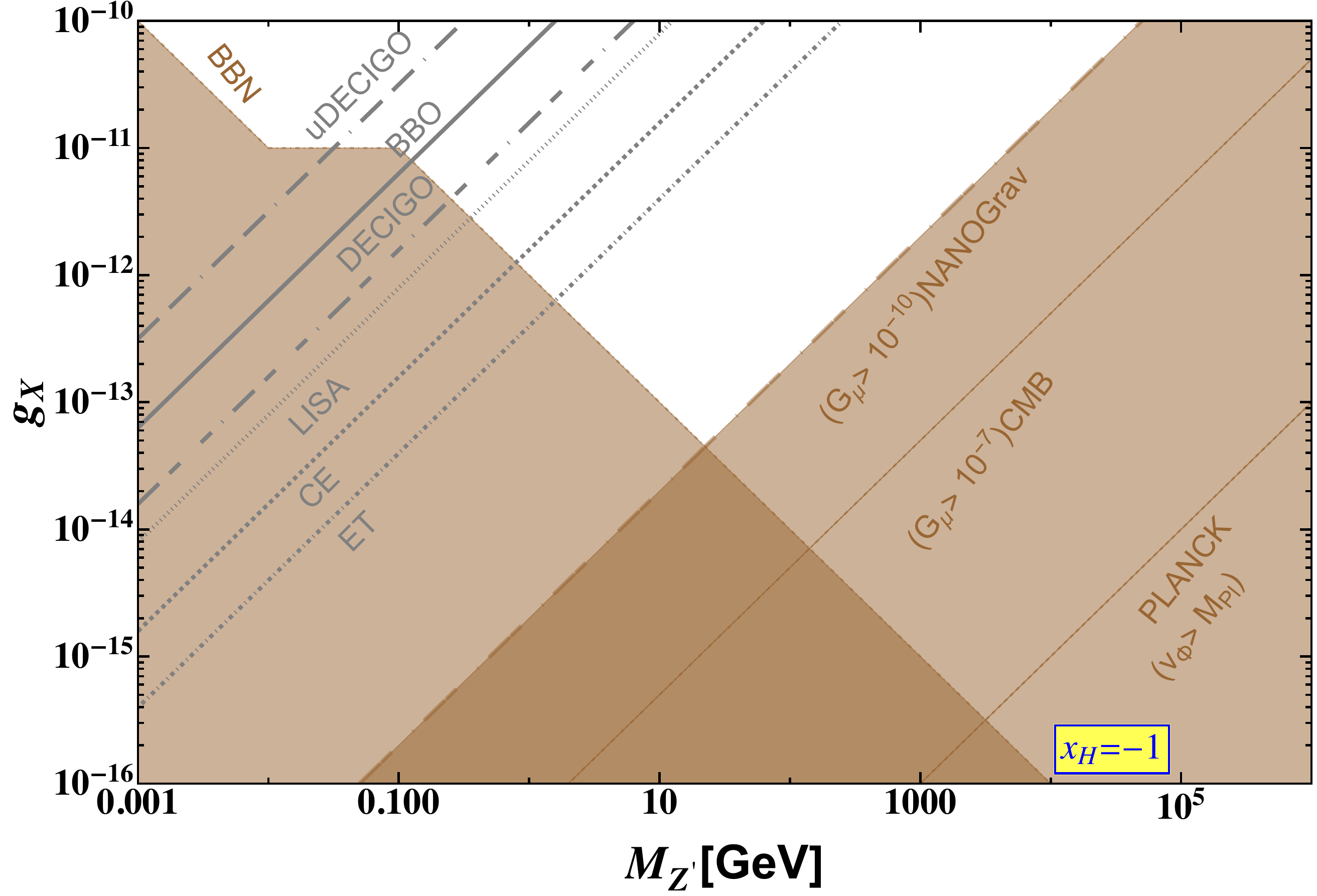}
\includegraphics[width=0.4973\textwidth,angle=0]{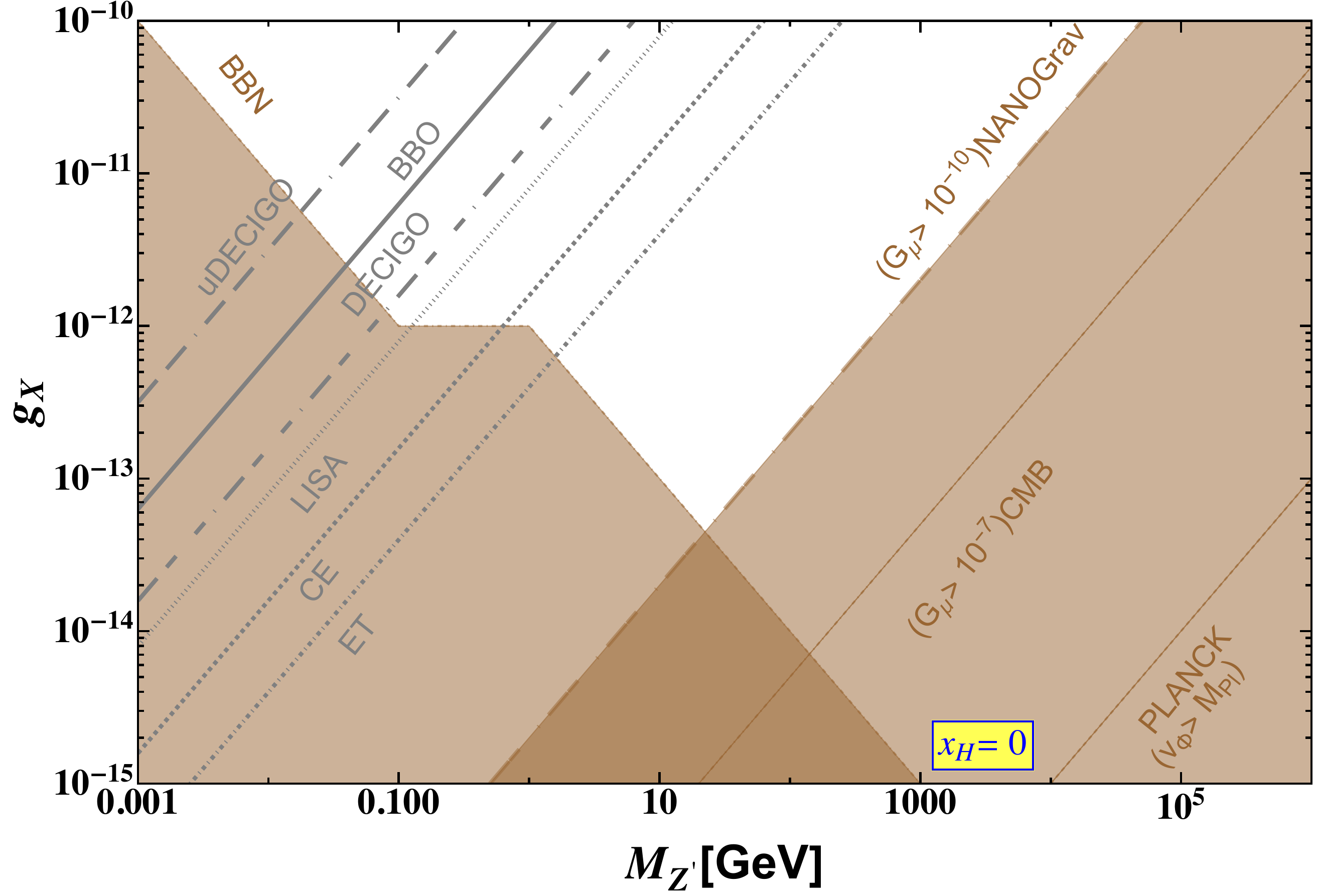}
\includegraphics[width=0.4973\textwidth,angle=0]{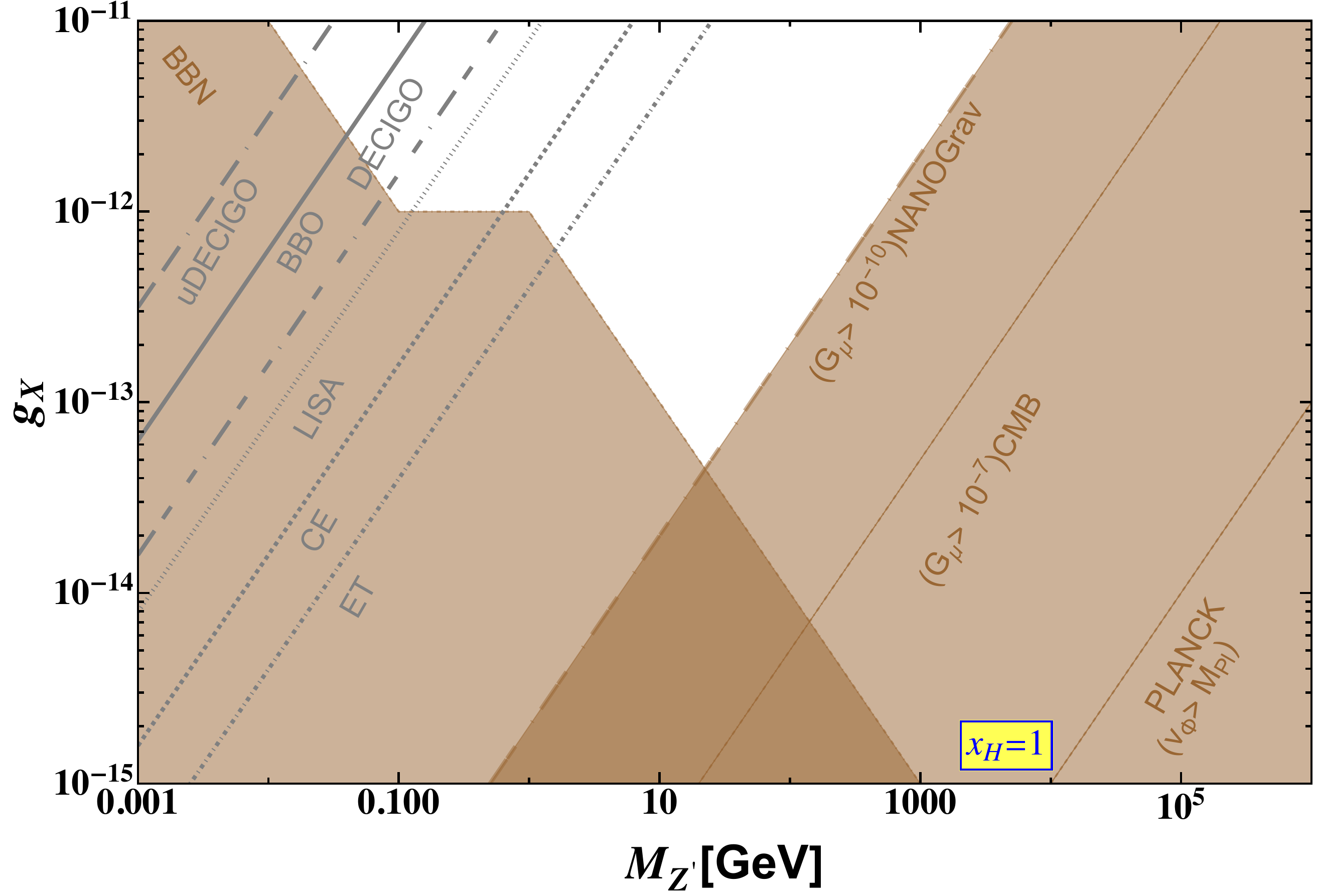}
\includegraphics[width=0.4973\textwidth,angle=0]{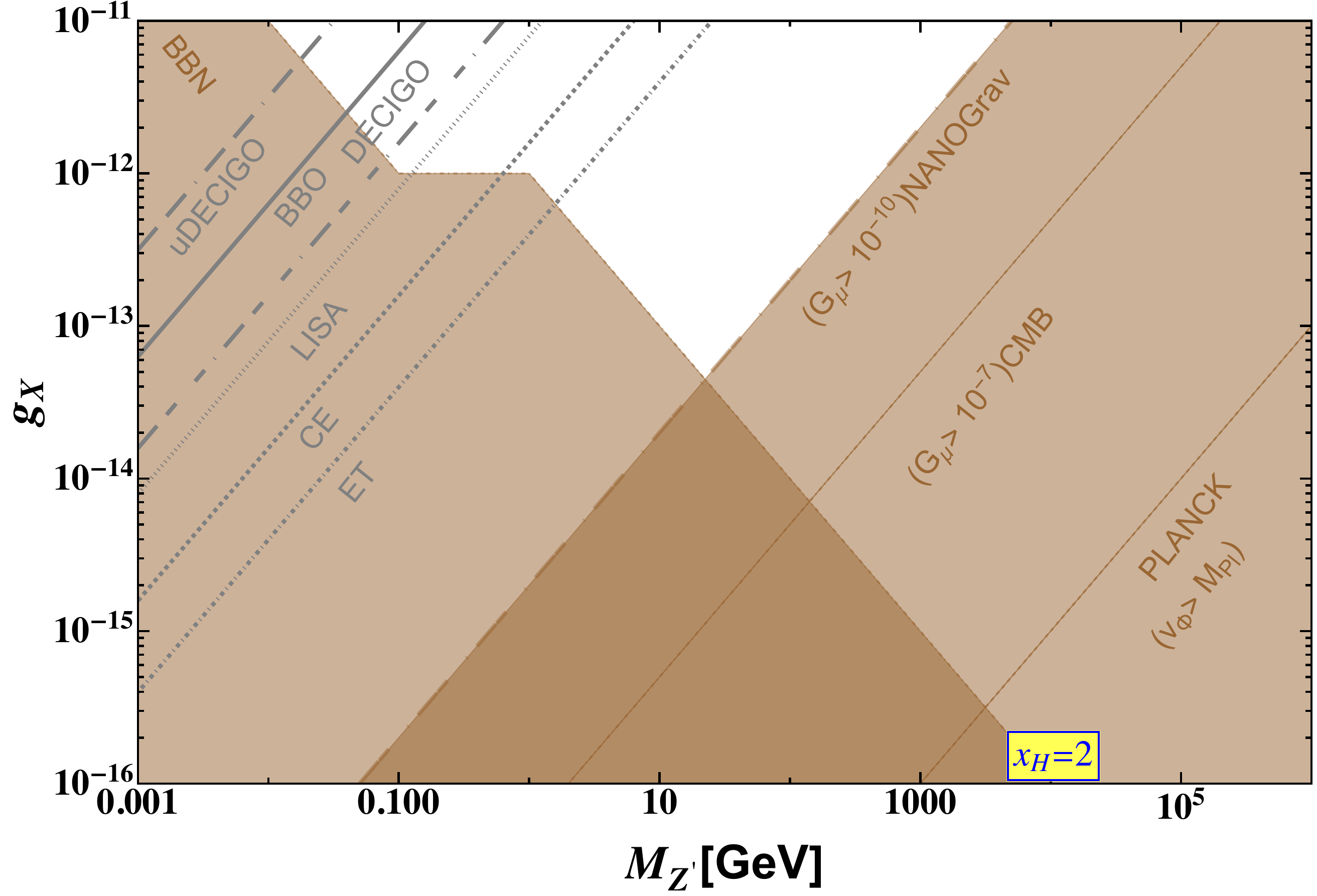}
\caption{Elaborated limits on $g_X-M_{Z^\prime}$ plane from current (NANOGrav) and future (uDECIGO, BBO, DECIGO, LISA, CE, ET) GW search experiments where GW is considered to be produced from cosmic strings for different $x_H$. We compare our results with the bounds obtained from BBN and CMB data. Bounds obtained from the condition $v_\Phi > 10^{19}$ GeV is represented as PLANCK. Shaded regions are ruled out by existing experimental bounds.}
\label{fig:summary2}
\end{figure*}
%%%%%%%%%%%%%%%%%%%%%%%%%%%%%%%%%%%%%%%%%%%%%%%%%%%%%%%%%%%%%%%%%%%%%%%%%

\noindent
{\textbf{Freeze-in production of dark matter}--}
%%%%%%%%%%%%%%%%%%%
%In this section, we derive constraints on the coupling strength and mass parameters by requiring the correct dark matter (DM) relic abundance. 
We focus on scenarios where the DM–SM interaction is feebler than the weak interaction, such that DM never thermalizes in the early Universe. This naturally leads to DM production via the \emph{freeze-in} mechanism, where the DM yield typically saturates when $T=\text{max}\left[m_\chi,\,\Mzp\right]$. Here, DM is produced dominantly through two processes:  (a) on-shell 1-to-2 decays of the $Z'$ gauge boson into DM pairs, and  
(b) $Z'$-mediated 2-to-2 $s$-channel scattering of thermal bath particles into DM pairs. We kinematically forbid the former channel by setting $m_\chi = 3\,M_{Z'}$. The evolution of the DM number density is governed by the Boltzmann equation (BEQ), which can be expressed in terms of the DM yield $Y_{\rm DM} \equiv n_{\rm DM}/s$, with $s$ being the entropy density. In terms of the dimensionless parameter $x \equiv m_\chi/T$, the equation reads,
\begin{equation}\label{eq:beq}
x\,H\,s\,\frac{dY_{\rm DM}}{dx} = \gamma_{22}\,,
\end{equation}
where $\gamma_{22}$ is the reaction density for 2-to-2 processes. Explicitly, this is given by:
\begin{align}
& \gamma_{22}=\frac{T}{32\pi^4}\,g_a g_b \int_{\mathfrak{s}_{\rm min}}^\infty ds\,\frac{\left[(\mathfrak{s} - m_a^2 - m_b^2)^2 - 4m_a^2 m_b^2\right]}{\sqrt{s}}
\nonumber \\&
\sigma(s)_{a,b\to1,2}\,K_1\left(\frac{\sqrt{\mathfrak{s}}}{T}\right), \label{eq:gam-ann}
\end{align}
where $g_{a,b}$ are the degrees of freedom of the initial states $a, b$ and $\mathfrak{s}_{\rm min} = \max\left[(m_a + m_b)^2, (m_1 + m_2)^2\right]$. During radiation domination, the Hubble rate and entropy density are given by,
\begin{align}
H(T) &= \frac{\pi}{3}\sqrt{\frac{\gs(T)}{10}}\,\frac{T^2}{M_P}, &
s(T) &= \frac{2\pi^2}{45}\,\gss(T)\,T^3,
\end{align}
where $\gs$ and $\gss$ are the relativistic degrees of freedom contributing to energy and entropy density, respectively.

To determine whether DM production is truly non-thermal, we compare the scattering rate $\Gamma_{22} = \gamma_{22}/n_{\rm eq}$ with the Hubble rate. The equilibrium number density of DM is given by $n_{\rm eq}^{\rm DM} = \left(T/(2\pi^2)\right)\,m_\chi^2\,K_2(m_\chi/T)$.
For an analytical estimate, we work in the limit $T \gg m_\chi, M_{Z'}$. This gives
\begin{align}\label{eq:rate}
&\frac{\Gamma_{22}}{H}\Bigg|_{T = M_{Z'}} \simeq \mathcal{C} \left(\frac{g_X}{10^{-4}}\right)^4 \left(\frac{M_{Z'}}{1\,\text{TeV}}\right)^2 \left(\frac{Q_\chi}{100}\right)^2\,,
\end{align}
for $m_\chi = 3\,M_{Z'}$ using the cross-sections from~\cite{Barman:2024lxy,Barman:2025bir} where $\mathcal{C}=\{3\times 10^{-5}, 10^{-4}, 3\times 10^{-4}, 6\times 10^{-4}\}$ for $x_H=\{-1,0,1,2\}$ and $x_\Phi=1$. This is a conservative estimate, as it uses the equilibrium number density for the DM. Since $\sigma(s) \propto g_X^4$, the interaction rate falls below the Hubble rate for relatively larger values of $g_X$, however, to ensure out of equilibrium production, we conservatively take $g_X \lesssim 10^{-3}$. The observed DM relic abundance requires $Y_0\, m_\chi = \Omega_{\rm DM}\,h^2\,\frac{\rho_c}{s_0\,h^2} \simeq 4.3 \times 10^{-10}\,\text{GeV}$, where $Y_0 = Y_{\rm DM}(T_0)$ is the present-day DM yield. Here $\rho_c \simeq 1.05 \times 10^{-5}\,h^2\,\text{GeV/cm}^3$, $s_0 \simeq 2.69 \times 10^3\,\text{cm}^{-3}$ and $\Omega_{\rm DM} h^2 \simeq 0.12$ are critical density, current entropy density \cite{ParticleDataGroup:2022pth} and observed relic density \cite{Planck:2018vyg}. In the high-temperature limit $T\gg m$, the asymptotic yield approximately follows $Y_0\propto g_X^4\, Q_\chi^2/\Mzp$, whereas, for $T\ll m$, the yield scales as $Y_0 \propto g_X^4\, Q_\chi^2/\Mzp^2$. Thus, in the heavy mass regime, achieving the same yield necessitates a larger coupling.  
\\
    
%%%%%%%%%%%%%%%%%%%%%%%%%%%
\noindent
{\textbf{Gravitational waves from cosmic string}--}
%%%%%%%%%%%%%%%%%%%%%%%%%%%%%%%%%%%
The dominant mechanism for energy loss from cosmic strings is GW emission from oscillating loops, as shown in simulations using the Nambu-Goto action~\cite{Ringeval:2005kr,Blanco-Pillado:2011egf}. The power radiated is given by~\cite{Vilenkin:1981bx}
\begin{equation}
    P_{\rm GW} = \frac{G}{5} (\dddot{Q})^2 \propto G\mu^2\,,
\end{equation}
where $\mu$ is the string tension. Thus, the energy loss rate reads,
\begin{equation}
    \frac{dE}{dt} = -\Gamma G\mu^2\,,
\end{equation}
with $\Gamma \simeq 50$~\cite{Vachaspati:1984gt}. The loop length evolves from its initial size $l_i = \alpha t_i$ as,
\begin{equation}
    l(t) = \alpha t_i - \Gamma G\mu(t - t_i)\,,
\end{equation}
where $\alpha \sim 0.1$ is the loop size parameter~\cite{Blanco-Pillado:2013qja,Blanco-Pillado:2017oxo}. Loops emit GWs in discrete harmonics with frequencies
\begin{equation}
    f_k = \frac{2k}{l(t)}\,, \quad k = 1,2,3,\dots\,.
\end{equation}
The present-day GW spectrum is defined as,
\begin{equation}
    \Omega_{\rm GW}(t_0,f) = \sum_k \Omega_{\rm GW}^{(k)}(t_0,f) = \frac{f}{\rho_c} \frac{d\rho_{\rm GW}}{df}\,.
\label{eqn:omgcs1}
\end{equation}
Since GW energy redshifts as $a^{-4}$, the contribution from each mode is~\cite{Blanco-Pillado:2013qja}
\begin{equation}
    \frac{d\rho_{\rm GW}^{(k)}}{df} = \int_{t_F}^{t_0} \left[\frac{a(t_E)}{a(t_0)}\right]^4 P_{\rm GW}(t_E,f_k) \frac{dF}{df} dt_E\,,
\label{eqn:omgcs2}
\end{equation}
where $dF/df = f[a(t_0)/a(t_E)]$. The power emitted per frequency is
\begin{equation}
    P_{\rm GW}(t_E,f_k) = \frac{2k G\mu^2 \Gamma_k}{f[a(t_0)/a(t_E)]^2}\,n\left(t_E, \frac{2k}{f}\left[\frac{a(t_E)}{a(t_0)}\right]\right)\,,
\label{eqn:omgcs3}
\end{equation}
with
\begin{equation}
    \Gamma_k = \frac{\Gamma\,k^{-4/3}}{\sum_{m=1}^\infty m^{-4/3}}\,, \quad \sum_k \Gamma_k = \Gamma\,.
\end{equation}
The loop number density $n(t_E,l)$ depends on the background cosmology $a(t) \propto t^\beta$ and is given by~\cite{Martins:1996jp,Martins:2000cs,Auclair:2019wcv}
\begin{equation}
    n(t_E,l) = \frac{A_\beta}{\alpha} \frac{(\alpha + \Gamma G\mu)^{3(1-\beta)}}{[l + \Gamma G\mu t_E]^{4 - 3\beta} t_E^{3\beta}}\,,
\label{eqn:omgcs4}
\end{equation}
where $A_\beta$ is a constant. Assuming cusp-dominated emission~\cite{Damour:2001bk,Gouttenoire:2019kij}, the present-day GW spectrum becomes
\begin{equation}
    \Omega_{\rm GW}^{(k)}(t_0,f) = \frac{2k G\mu^2 \Gamma_k}{f \rho_c} \int_{t_{\rm osc}}^{t_0} dt \left[\frac{a(t)}{a(t_0)}\right]^5 n(t,l_k)\,,
\label{eqn:omgcsfin}
\end{equation}
where $t_{\rm osc}$ marks the end of the friction-dominated era~\cite{Vilenkin:1991zk}. During radiation domination, the spectrum features a flat plateau with amplitude
\begin{equation}
    \Omega_{\rm GW}^{(k=1),{\rm plateau}}(f) = \frac{128\pi G\mu}{9\zeta(4/3)} \frac{A_r}{\epsilon_r} \Omega_r \left[(1 + \epsilon_r)^{3/2} - 1\right],
\end{equation}
where $\epsilon_r = \alpha / (\Gamma G\mu)$, and $A_r = 0.54$ for the radiation era~\cite{Auclair:2019wcv}. 
\\

%%%%%%%%%%%%%%%%%%%%
\noindent
{\textbf{Results and discussions}--}
We derive our limits by examining nearby NS and BD. The most constraining bounds primarily arise from three factors: (i) small distance to the earth ($d_{\star}$), (ii) strong gravitational potential ($M_{\star}/R_{\star}$), and (iii) comparatively old stellar age ($t_{\star}$). Guided by these considerations, for BDs, we select the target WISE J104915.57–531906.1~\cite{Burgasser_2013}, characterized by $d_{\star} \sim 2$ pc, $M_{\star} = 33.5~M_{\rm jupiter}$, $R_{\star} = 0.85~R_{\rm jupiter}$, and $t_{\star} = 4.5$ Gyr. For NSs, we consider PSR J0711–6830~\cite{Bramante:2024ikc} with $d_{\star} \sim 110$ pc, $M_{\star}$ = 1.4 $M_{\odot}$, $R_{\star} = 10$ km, and $t_{\star} = 5.84$ Gyr~\cite{atnf}. We then conservatively compare the expected flux from Eq.~\eqref{eq:exp_fluc} with the projected differential flux sensitivities for track-like events at declination $\delta = 0^\circ$, where detectors achieve their best point-source performance, using IC1~\cite{IceCube:2019cia}, KM3~\cite{KM3NeT:2018wnd}, the proposed IIC2~\cite{IceCube-Gen2:2021tmd}, and TD~\cite{TRIDENT:2022hql}. This choice corresponds to the most constraining flux limits currently available, and we derive the resulting bounds on the $[g_X, M_{Z^\prime}]$ parameter space\footnote{Declination-dependent sensitivities are typically reported by neutrino telescopes, with the most constraining limits obtained near the celestial equator ($\delta \simeq 0^\circ$); 
see, e.g., \cite{IceCube:2019cia, KM3NeT:2018wnd}. We adopt this benchmark as an optimistic reference, noting that sensitivities at other declinations or for extended sources would be weaker.}. For the $2\nu$ channel, bounds from NSs (dot--dashed) are stronger than those from BDs (dashed), as shown in Fig.~\ref{fig:summary1} for IC1 (red), KM3 (magenta), IC2( blue) and TD (purple) detectors. For comparison, we also include freeze-in limits consistent with the observed relic abundance (solid orange lines), discussed below Eq.~\eqref{eq:rate}. Notably, the IC1 bound intersects the freeze-in curve near $M_{Z^\prime}\simeq 0.6~\text{GeV}$ and weakens slightly around $M_{Z^\prime}\simeq 10~\text{GeV}$. The KM3 bound is marginally stronger than the freeze-in line, becoming nearly coincident near $M_{Z^\prime}\simeq 1~\text{GeV}$ for larger $U(1)_X$ charges, e.g., $x_H=2$. Prospective sensitivities from IC2 and TD surpass both the freeze-in and current experimental bounds. For the $4\nu$ channel, we evaluate constraints from long-lived $Z^\prime$ pairs produced in DM annihilation. The corresponding bounds, shown as solid curved lines in Fig.~\ref{fig:summary1}, cover the region $10^{-3}\,\text{GeV} \leq M_{Z^\prime} \leq 10^{-2}\,\text{GeV}$ and $10^{-9}\leq g_X\leq 10^{-6}$ for different $U(1)_X$ charges, derived from IC1 (red), KM3 (magenta), IC2 (blue), and TD (purple) considering $m_{\chi}=3 M_{Z^\prime}$. We also showed existing bounds considering $m_\chi=100 M_{Z^\prime}$ in black for IC1 (dashed), KM3 (dot-dashed), IC2 (dotted) and TD (solid). With increase in $x_H$ bounds become weaker. In case of $x_H=2$, IC1 and KM3 lines seem to coincide due to small difference between them for $m_\chi=100\,M_{Z^\prime}$.

Stringent constraints on $[g_X,\,M_{Z^\prime}]$ plane have already been established from long-lived $Z^\prime$ searches at proton beam-dump experiments such as NOMAD~\cite{NOMAD:2001eyx}, $\nu$-cal~\cite{Blumlein:1990ay,Barabash:2002zd}, CHARM~\cite{CHARM:1985nku}, FASER~\cite{FASER:2023tle,FASER:2024bbl}, NA62~\cite{NA62:2025yzs}, LSND~\cite{LSND:1997vqj,LSND:2001aii}, and PS191~\cite{Bernardi:1985ny}, as well as electron beam-dump experiments including E141~\cite{Riordan:1987aw}, E137~\cite{Bjorken:1988as}, E774~\cite{Bross:1989mp}, Orsay~\cite{Davier:1989wz} and KEK~\cite{Konaka:1986cb}. These searches exclude a wide region of parameter space for $\Mzp\lesssim 1$~GeV, as shown in Fig.~\ref{fig:summary1}. In addition, we also indicate bounds from low-energy scattering experiments \cite{Asai:2023mzl}, including dark photon searches at LHCb~\cite{LHCb:2019vmc}, neutrino--electron scattering in BOREXINO~\cite{Borexino:2000uvj,Bellini:2011rx}, TEXONO~\cite{TEXONO:2009knm}, GEMMA~\cite{Beda:2010hk}, CHARM-II~\cite{CHARM-II:1993phx,CHARM-II:1994dzw}, and neutrino--nucleon scattering from COHERENT~\cite{COHERENT:2018imc,COHERENT:2020iec,COHERENT:2020ybo}. 
% \textcolor{red}{We show bounds from supernova cooling by orange dashed line \cite{Asai:2022zxw} which are partly comparable \bb{??} with the $4\nu$ bound from long-lived $Z^\prime$ taking NSs in account when $m_\chi=3 M_{Z^\prime}$ and the respective bounds become weaker when $m_\chi=100 M_{Z^\prime}$. For $x_H=-1$, this scenario could provide a new stronger indirect bound $g_X\simeq 3\times 10^{-8}$ for $0.02$ GeV $\leq M_{Z^\prime} \leq 0.045$ GeV compared to previous beam-dump experiments. Parameters shown in Figs.~\ref{fig:summary1} and \ref{fig:summary2} are safe from purterbatibity constraints $Q_\chi g_X < \sqrt{4\pi}$.} 
We show bounds from supernova cooling by orange dashed lines. These bounds are partially comparable to the $4\nu$ constraints from long-lived $Z^\prime$ production in neutron stars when $m_\chi = 3 M_{Z^\prime}$, while they become weaker for $m_\chi = 100 M_{Z^\prime}$. For $x_H = -1$, this scenario can yield a new, stronger indirect bound of $g_X \simeq 3\times10^{-8}$ in the mass range $0.02~\text{GeV} \leq M_{Z^\prime} \leq 0.045~\text{GeV}$, improving upon previous beam-dump limits. On the other hand, for BDs, $P_{\rm surv}$ remains nonzero for $\Mzp < 1~\mathrm{MeV}$, well below $m_{\rm evap} \simeq 0.7~\mathrm{GeV}$, so in the allowed mass range the expected flux lies several orders of magnitude below the detector sensitivity, preventing limits in the $4\nu$ channel. The parameter space shown in Figs.~\ref{fig:summary1} and \ref{fig:summary2} also satisfies the perturbativity condition $Q_\chi g_X < \sqrt{4\pi}$.

In heavier $\Mzp$ regime, the diagonal gray dashed lines in indicate values of $v_\Phi$ that lie within the sensitivity reach of proposed gravitational-wave detectors: Big Bang Observer (BBO)~\cite{Crowder:2005nr,Corbin:2005ny}, ultimate DECIGO (uDECIGO)~\cite{Seto:2001qf,Kudoh:2005as}, LISA~\cite{LISA:2017pwj}, the Cosmic Explorer (CE)~\cite{Reitze:2019iox}, and the Einstein Telescope (ET)~\cite{Hild:2010id,Punturo:2010zz,Sathyaprakash:2012jk,Maggiore:2019uih}. We extend these bounds in Fig.~\ref{fig:summary2}, where the light-brown shaded region in the lower-left corner corresponds to lifetime $\tau_{Z^\prime}\equiv 1/\Gamma_{Z^\prime} > 1~\text{sec}$, thereby conflicting with big bang nucleosynthesis (BBN) predictions. The dark-brown shaded regions in the lower-right corner are excluded by multiple constraints: CMB observations requiring $G\mu \lesssim 10^{-7}$~\cite{Charnock:2016nzm}, 
recent NANOGrav results demanding $G\mu \lesssim 10^{-10}$~\cite{NANOGrav:2023hvm}, 
as well as the theoretical upper bound excluding super-Planckian values $v_\Phi > M_P$. 
\\

%%%%%%%%%%%%%%%%%%%%%%%%%%
\noindent
{\textbf{Conclusions}--}
%%%%%%%%%%%%%%%%%%%%%%%%
Dark matter capture in astrophysical objects offers powerful probes of new physics, complementary to experimental searches like beam-dump, low energy scattering and cosmological observations. In this work, within a minimal UV-complete framework, we consider DM captured by neutron stars and brown dwarfs that annihilates via a massive gauge boson. We constrain the relevant parameter space by combining limits from capture rates, freeze-in relic abundance consistent with Planck data, gravitational waves from cosmic strings, big bang nucleosynthesis (BBN), and energy- and intensity-frontier experiments. Remarkably, we find that constraints obtained from NS and BD can test regions of  parameter space for $\Mzp\gtrsim 1$ GeV and $g_X\sim\left[10^{-6}-10^{-7}\right]$, consistent with right DM abundance, while inaccessible to GW detectors and several low and high energy experiments.
\\

%%%%%%%%%%%%%%%%%%%%%%%%%%
\noindent
{\textbf{Acknowledgments}--}
%%%%%%%%%%%%%%%%%%%%%%%%
We thank Sanjoy Mondal for fruitful discussions. PB acknowledges support from the COFUND action of Horizon Europe’s Marie Sklodowska-Curie Actions research programme, Grant Agreement 101081355 (SMASH).
%%%%%%%%%%%%%%%%%%%%%%%%%%%
%\end{widetext}
\vspace{-0.198in}
\bibliographystyle{utphys}
\bibliography{bibliography}
%%%%%%%%%%%%%%%%%%%%%%%%%%%
%%%%%%%%%%%%%%%%%%%%%%%%%%%
\end{document}